\documentclass[ag]{copernicus}

\usepackage{mathptmx}

\usepackage{graphicx} 

\usepackage{amsfonts}
\usepackage{amssymb}
\usepackage{amsbsy}
\usepackage[figuresright]{rotating}
\usepackage{multicol}
\usepackage{array}
\usepackage{supertabular}

\usepackage{color}

\begin{document}

\title{{Collisionless reconnection: The sub-microscale mechanism \\ of magnetic field line interaction }}

\author[1,2]{{R. A. Treumann}%\thanks{Visiting the International Space Science Institute, Bern, Switzerland}
}
\author[3]{{W. Baumjohann}}
\author[4]{{W. D. Gonzalez}}

\affil[1]{Department of Geophysics and Environmental Sciences, Munich University, Munich, Germany}
\affil[2]{Department of Physics and Astronomy, Dartmouth College, Hanover NH 03755, USA}
%\affil[3]{International Space Science Institute, Bern, Switzerland}
\affil[3]{Space Research Institute, Austrian Academy of Sciences, Graz, Austria}
\affil[4]{Divis$\tilde\mathrm{a}$o de Geof'sica Espacial, Instituto Nacional de Pesquisas Espaciais,  S‹o Jos\'e dos Campos, S$\tilde\mathrm{a}$o Paulo, Brazil}

\runningtitle{Collisionless Reconnection}

\runningauthor{R. A. Treumann, W. Baumjohann and W. D. Gonzales}

\correspondence{R. A.Treumann\\ (rudolf.treumann@geophysik.uni-muenchen.de)}

\received{ }
%\pubdiscuss{ } %% only important for two-stage journals
\revised{ }
\accepted{ }
\published{ }

%% These dates will be inserted by the Publication Production Office during the typesetting process.

\firstpage{1}

\maketitle

\begin{abstract}
Magnetic field lines are quantum objects carrying one quantum $\Phi_0=2\pi\hbar/e$ of magnetic flux and have finite radius $\lambda_m$. Here we argue that they possess a very specific dynamical interaction. Parallel field lines reject each other. When confined to a certain area they form two-dimensional lattices of hexagonal structure. We estimate the filling factor of such an area. Antiparallel field lines, on the other hand, attract each other. We identify the physical mechanism as being due to the action of the gauge potential field which we determine quantum mechanically for two parallel and two antiparallel field lines. The distortion of the quantum electrodynamic vacuum causes a cloud of virtual pairs. We calculate the virtual pair production rate from quantum electrodynamics and estimate the virtual pair cloud density, pair current and Lorentz force density acting on the field lines via the pair cloud. These properties of field line dynamics become important in collisionless reconnection, consistently explaining why and how reconnection can spontaneously set on in the field-free centre of a current sheet below the electron-inertial scale.

\end{abstract}

\introduction
The concept of magnetic field lines is central to plasma physics. They can be `frozen-in' to the plasma becoming transported with the plasma flow if only dissipative processes are negligible. Space plasma physics has made wide application of these concepts in the large-scale behaviour of collisionless plasmas. Problems arise when small-scale processes are observed, magnetic fields cross narrow boundaries and `reconnect' with field lines of opposite or inclined orientation, processes which in a collisionless plasma are hard to understand. In such cases the field-line concept offers some simple geometrical scenarios of cutting and merging but raises the question of the identity and nature of magnetic field lines. 

Based on well-established quantum mechanical principles following the seminal work of \citet{aharonov1959} and \citet{landau1930}, we have demonstrated in a previous communication \citep{treu2011} that magnetic field lines can be understood as magnetic flux quanta occupying a flux tube of a certain well-defined cross section in a given magnetic field $\mathbf{B}$. Single field line merging is understood as the annihilation of two such magnetic flux quanta in the contact of two strictly oppositely directed field line sections over the length $\ell_\|$ along the field lines. 

The micro-scale field-line merging and microscopic flux quantum annihilation raises the non-trivial question of how magnetic field lines can be brought into mutual contact. This question is part of the more general problem of the mechanism of interaction between magnetic field lines. The present communication is devoted to its investigation as a pre-requisite to the understanding of macro-scale reconnection.

\section{The field-line concept}
Magnetic field lines carry single flux quanta $\Psi_0=\pm\Phi_0=\pm2\pi\hbar/e$, defined by elementary constants of nature, the quantum of action $\hbar$ and the elementary charge $e$. Thus, $\Phi_0$ itself is a constant of nature.\footnote{Note that $e$ is the renormalised charge in quantum field theory.} The existence of magnetic flux quanta was experimentally confirmed by \citet{klitz1980} spectacular discovery of the Quantum Hall Effect. The flux can be positive or negative, depending on the direction of the magnetic field $\pm\mathbf{B}$. Each magnetic field line being a flux tube of radius\begin{equation}
\lambda_m=\sqrt{\frac{\Phi_0}{\pi B}}
\end{equation}
the `magnetic length' of \citet{landau1930}.  
He inferred this length from the quantum-mechanical motion of an electron in a homogeneous magnetic field, finding that the perpendicular electron energy $\epsilon_\perp=\hbar\omega_{ce}(n+{\frac{1}{2}})$ would be quantised, with $\omega_{ce}=eB/m_e$ the non-relativistic\footnote{Having in mind application to space problems like reconnection at the magnetopause and in the geomagnetic tail, we do not attempt to treat the relativistic case here.} electron cyclotron frequency, $m_e$ electron rest mass, and $n=0, 1, ...$. One easily recognises that the `magnetic length' corresponds to the gyroradius of an electron in the lowest Landau level (LLL), i.e. the gyroradius of an electron of very low energy. Its coincidence with the field-line radius implies that the \emph{smallest gyro-cross section} an electron could possess in a given magnetic field $B$ is the cross -section of a magnetic flux tube of one quantum of magnetic flux $\Phi_0$, i.e. one field-line. Under space conditions near Earth with magnetic fields of the order of $1\lesssim B\lesssim 10^5$ nT,  this energy is very small,$10^{-13}\lesssim\epsilon_\mathrm{\,LLL}\lesssim10^{-8}$ eV. At electron temperatures of the order of 1 eV $\lesssim T_e$ in space, the low Landau levels will be empty. Thermal Landau levels (TLL) occupied by thermal electrons have quantum numbers centred around $n_\mathrm{\, TLL}\gtrsim 10^{10}$ and forming thermal continua. 

Hence, one distinguishes between the respective low energy and thermal energy regimes. The former is the strongly coupled regime of the Quantum Hall Effect.  Strong coupling means that electrons and magnetic field quanta are closely tied via the Laughlin wave function of the electrons \citep{laughlin1983} which extends the Landau solution to many electrons. In this low-energy regime, electrons are forced to occupy Landau levels.  Electrons in the lowest Landau level plus flux quanta form a fluid consisting of quasi-Fermions or `composite electrons' with effective charge $q=e/3$ and, for particular magnetic field strengths, lead to quantisation of the Hall resistance.

In the thermal regime, electrons become independent of flux quanta. Their dynamic scales, being of the order of the electron gyroradius $r_{ce}\gg\lambda_m$, grossly exceed the field-line scale, and the coupling between electrons and field lines becomes weak. On the other hand, on scales $<r_{ce}$ below the electron gyroradius the dynamics of magnetic flux quanta is independent of the presence of electrons. While being adiabatically enslaved to large numbers $N_\mathit{ce}$ by the gyrating electrons, flux quanta may undergo mutual short-range interaction, here. Enslavement is due to the centripetal Lorentz force of the gyrating electron which on the scale $\sim r_{ce}$ is balanced by the pressure gradient of the total number of magnetic field lines in the gyration cross section $\pi r_\mathit{ce}^2$. This number  is proportional to the ratio of the respective gyration to field line cross-sections
\begin{equation}
N_\mathit{ce}\sim \frac{\pi r_\mathit{ce}^2}{\pi\lambda_m^2}=\frac{\cal{E}_\perp}{\hbar\omega_{ce}}
\end{equation}
where ${\cal E}_\perp=p_\perp^2/2m_e$, and $p_\perp$ is the perpendicular electron momentum. This number is not completely conserved, however, during one electron gyration. Gyration takes time $\Delta t\sim 2\pi/\omega_\mathit{ce}$ during which a field line may escape to the environment. The uncertainty of $N_\mathit{ce}\approx \Delta{\cal E}_\perp/\hbar\omega_\mathit{ce}$ is obtained from the uncertainty relation $\Delta{\cal E}_\perp\Delta t\sim h$ to amount to only
\begin{equation}
\Delta N_\mathit{ce}\sim 1
\end{equation}
Hence, during one electron gyration the number of the many adiabatically trapped field lines in the electron gyration cross-section either lost or added to the frozen-in magnetic flux is of the order $\Delta N_\mathit{ce}=O(1)$ only. Under frozen-in conditions this number forms an electron gyration flux tube and is convected together with the electron across the plasma.

All these frozen-in magnetic field lines are parallel to each other within mutual inclination angles $0\leq\theta<\frac{1}{2}\pi$, being unable to undergo any merging  \citep{treu2011}. The angular deviations may be caused by magnetic fluctuations of various origin and are of no interest for the following. During convection of the plasma, the whole bunch of $N_\mathit{ce}$ frozen-in field lines is carried across the plasma. Referring to the reconnection site, it is carried toward the centre of the current sheet in the process of reconnection when the electrons cross the `ion diffusion region' but  for long have not yet entered the electron inertial region in the centre of the reconnection site.
\begin{figure}[t!]
\centerline{{\includegraphics[width=0.5\textwidth,clip=]{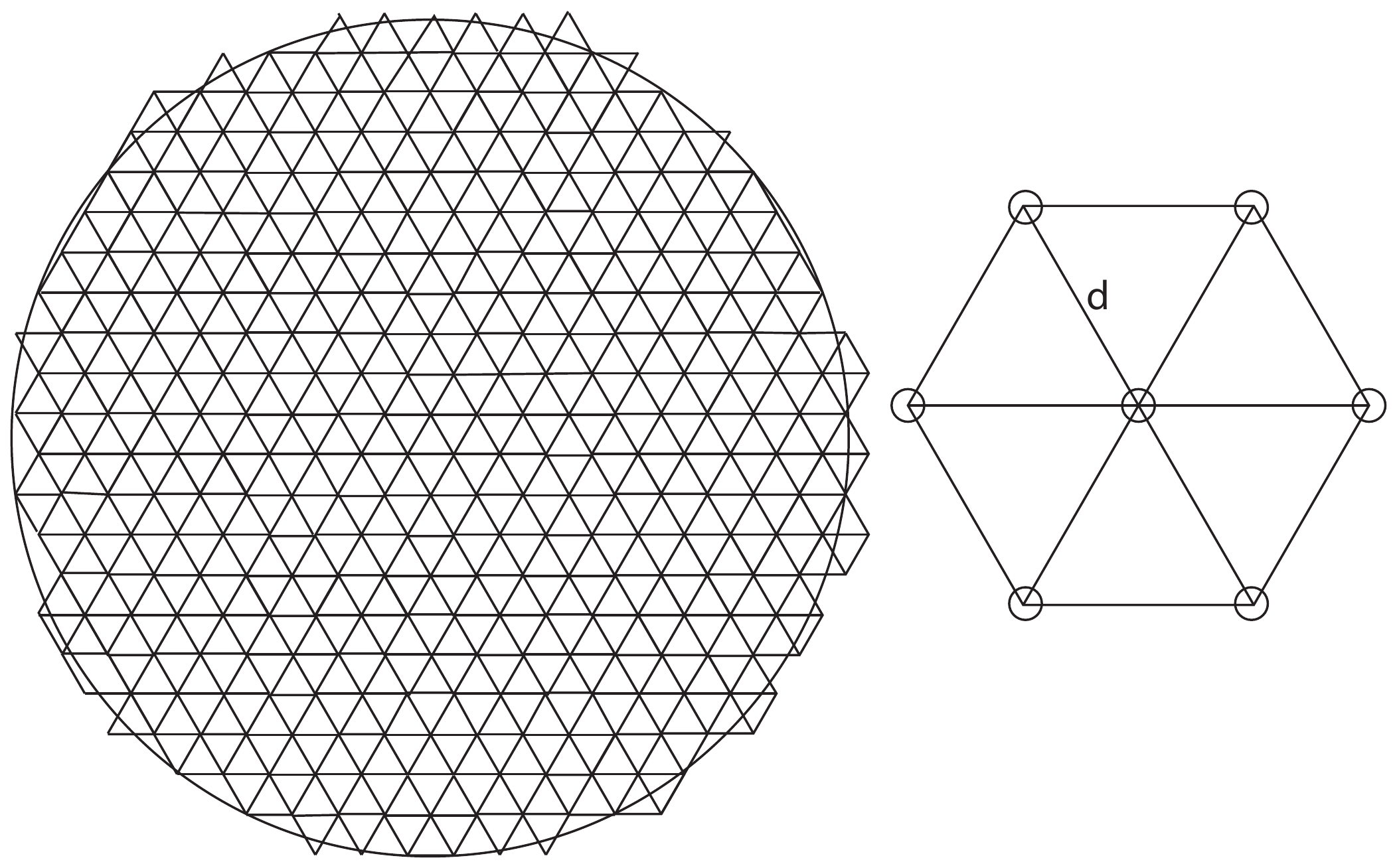}
}}
\caption[ ]
{\footnotesize {Repulsive interaction between confined parallel field lines generates a hexagonal lattice composed of elementary flux tubes, magnetic field lines, each carrying just one flux quantum $\Phi_0=2\pi\hbar/e$. Every corner of a single hexagon is occupied by a field line, as shown on the right. If field lines are trapped like in the frozen-in concept then the number of field lines is just six times the number of hexagons which can be fitted into the cross section.}}
\vspace{-0.3cm}\label{fig-hex}
\end{figure}

\section{Field-line topology}
The first interesting point is that the value of the above number $N_\mathit{ce}$ is slightly over-estimated by the naive assumption of dense packing of field lines. Quantisation of the flux distorts the continuous distribution of magnetic fields. Parallel field lines (flux tubes) exert repulsive forces on each other seeking to expand into space and separate as distant as possible from neighbouring field lines. This is due to the Lorentz force-like interaction between the flux tubes and is well known from classical field-line patterns like, for instance, those of dipolar or quadrupolar fields. Classically the Lorentz force on a magnetic flux tube with field strength $\mathbf{B}_1$ exerted by a neighbouring flux tube of field strength $\mathbf{B}_2$ is given by
$\mathbf{F}_{12}=-\nabla(\mathbf{B}_1\cdot\mathbf{B}_2)/2\mu_0+(\mathbf{B}_1\cdot\nabla)\mathbf{B}_2/\mu_0$. The second term accounts for the stresses produced by bending or twisting the flux tubes. In the absence of any bending only the first term survives. Clearly, since $B\sim 1/r^\alpha$, with $\alpha>0$ some power, parallel field lines are subject to positive (repulsive), anti-parallel field lines to negative (attractive ) forces. 

\begin{figure}[t!]
\centerline{{\includegraphics[width=0.3\textwidth,clip=]{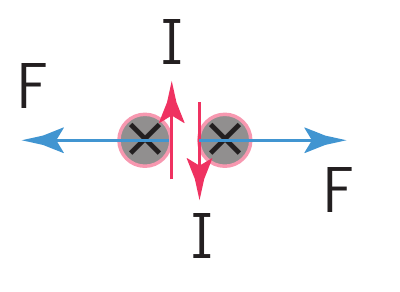}
}}
\caption[ ]
{\footnotesize {Repulsive interaction between two closely spaced parallel field lines pointing into the plane and each carrying one electron in its lowest Landau level, LLL. Electron gyration is clockwise and confined to the circumference of the field line (red circles), producing  anti-parallel currents $I$ red arrows). These currents give rise to repulsive forces $F$ (blue arrows).}}
\vspace{-0.3cm}\label{fig-lorf}
\end{figure}

This is illustrated in Figure \ref{fig-lorf} for the case of two closely spaced field lines which point into the plane. Putting one test-electron in LLL state on the circumference $R=r/\lambda_m=1$ of each field line, the electrons gyrate clockwise, each producing a circular line current $I=-I_0\delta(R-1)$, with $ I_0=e\sqrt{\hbar\omega_{ce}/m_e\lambda_m^2}\sim 10^{-21}\sqrt{\omega_{ce}/\lambda^2_m}$ A. The two anti-parallel currents, indicated by the red arrows, cause repulsive Lorentz forces between the parallel field lines of strengths per unit length $F\propto 2\mu_0I_0^2/r\sim 10^{-48}\sqrt{\omega_{ce}}/\lambda_m^3R$. The forces become attractive for two anti-parallel field lines where the currents flow parallel to each other. However, the field lines are located in vacuum, and no LLL electrons are available. In this case the interaction is not obvious as we will discuss in detail below in Sections 4 and 5.

For parallel field lines carrying one flux quantum only and being confined to a spatial cross-section like the gyration-cross section of the electron this subtle interaction causes a very particular arrangement of field-lines. Rejecting and keeping themselves at distances $d\gg\lambda_m$ from each other, in a homogeneous magnetic field they create a lattice of field lines consisting of hexagonal elements with one field line in the centre surrounded by six field lines in the corners of the hexagon. Every field line spanning its own hexagon with the space between the field lines being void of magnetic fields. This is shown in Figure \ref{fig-hex} and differs from the continuous field distribution in classical physics on the macro-scale. On the sub-microscale the quantisation of the magnetic flux causes a lattice structure of the magnetic field.

The homogeneous field-line filling factor follows from comparing the cross sections of field line and hexagon.  The latter consists of six triangles of side lengths $d'=d+2$ where $d'$ is the centre-to-centre distance between two neighbouring field lines, and $d >1$ is the external vertical distance between the field lines, both in units of $\lambda_m$. The field-line free surface of the hexagon (hexagon surface $S_H=3\sqrt{3}d'^2\lambda_m^2/2$ minus the surfaces $S_\mathit{fl}=3\pi\lambda_m^2$ contributed by the field line in the centre and those in the corners) becomes
\begin{equation}
S_\mathit{empty}=S_\mathit{fl}\left[2\sqrt{3}\left(\frac{d^2}{2}+1\right)-1\right]
\end{equation}
This yields the field-line filling factor
\begin{equation}
q_\mathit{fl}=\frac{S_\mathit{fl}}{S_H}=\frac{2\pi}{\sqrt{3}}(d+2)^{-2}
\end{equation}
The distance $d$ between the field lines is determined by the repulsive  force between the parallel field lines and is not easy to determine without any knowledge about the force between the field lines and the size of the volume of the plasma to which it is confined. We delay this question to the discussion in the next section.

Assuming that the field lines are confined to the electron gyro-cross section, the number of field lines in this cross section is determined dividing it by the surface of the hexagon, finding that an electron cyclotron-cross section contains 
\begin{equation}
N_\mathit{fl}^\mathit{ce}\sim \frac{2\pi N_\mathit{ce}}{\sqrt{3}(d+2)^2}
\end{equation}
field lines. This, as a result of the repulsive  force, is less by the filling factor than the originally given number $N_\mathit{ce}$. Because it depends only on the inter-field line distance $d$, the reduction factor holds for any arbitrary cross-section in homogeneous fields. Measuring an average (for instance over the electron cyclotron-cross section) magnetic field $\langle B\rangle$, the magnetic field $B$ of a single field line contained in the cross-section must be larger by the factor
\begin{equation}
\frac{B}{\langle B\rangle}\sim \frac{\sqrt{3}}{2\pi}(d+2)^2
\end{equation}

Unfortunately, there is no obvious and simple independent determination of the `lattice constant' $d$ because $d$ is a dynamical quantity which adjusts itself to Lorentz force equilibrium between the entire ensemble of field line flux tubes in the volume. Assuming that $B/\langle B\rangle\sim 10^3$ yields a lattice constant of $d\sim 15$. If $B/\langle B\rangle\sim 10$ this value reduces to $d\sim 3$ only. One may conclude that generally the field concentrated in one field line will be strong and the cross section small.

In a $\beta\sim 1$ plasma one has for the average magnetic field $\langle B\rangle^2\sim 2\mu_0N_eT_e$ obtaining for the magnetic field of one field line $B\sim \sqrt{3\mu_0N_eT_e/2}(d +2)^2$. This will hold approximately in the `ion diffusion region' until one enters the electron inertial range.

\section{Field-line interaction}

Classically there is no answer to the question of how the force is transmitted across the field-free space between the field lines. In the absence of LLL electrons, no field exists outside the field lines except for the gauge field $\mathbf{A}=\nabla\Lambda$ which does not directly contribute to any magnetic field.  It is, in fact,  the gauge field which takes care of the absence of magnetic fields outside the field line, keeping external space clean of magnetic fields. To use a common term: \emph{Field lines have no hair}.
\begin{figure}[t!]
\centerline{{\includegraphics[width=0.5\textwidth,clip=]{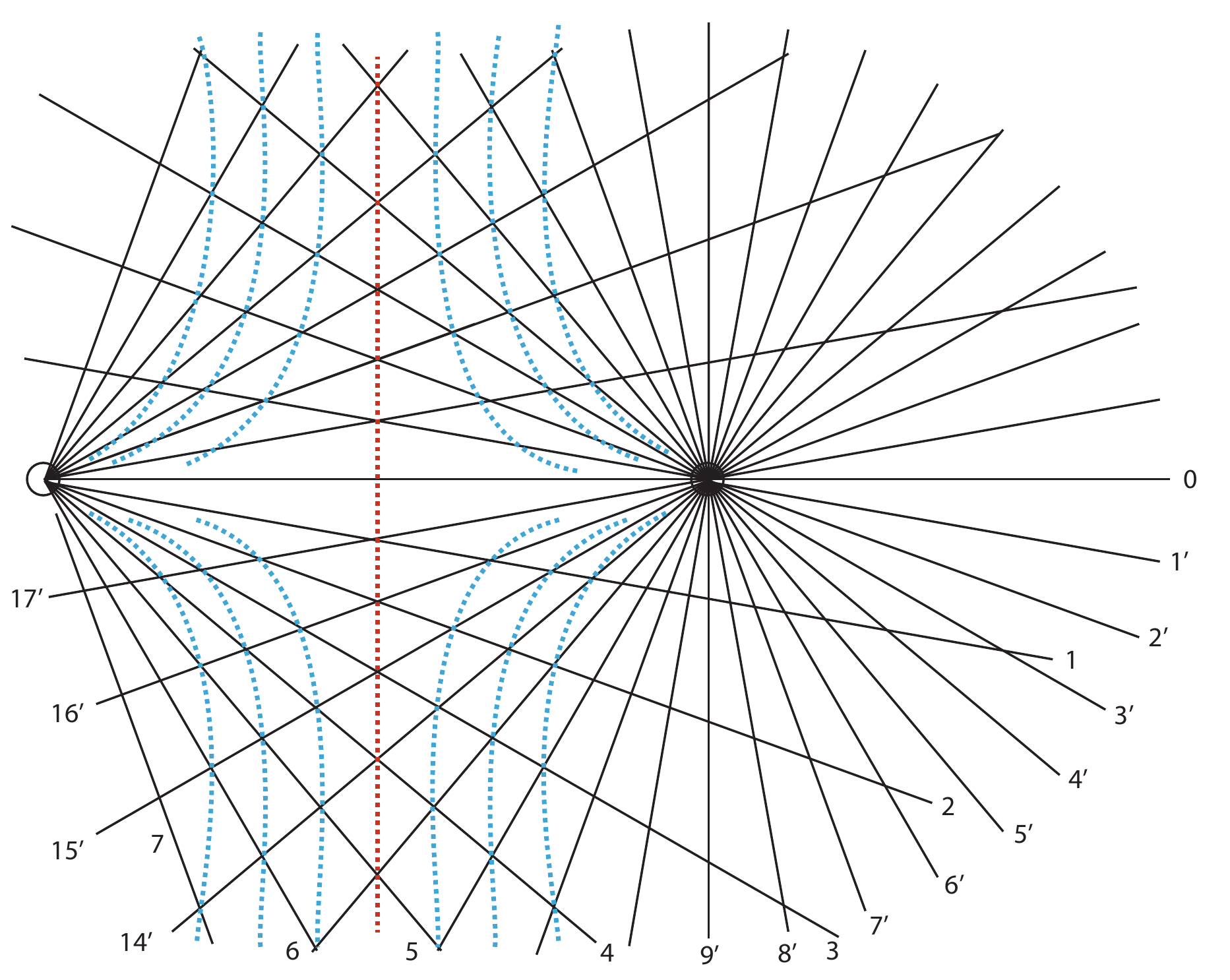}
}}
\caption[ ]
{\footnotesize {Superposition of the equi-gauge potential lines $\Lambda=$ const of two parallel magnetic field lines (elementary flux tubes carrying one flux quantum $\Phi_o=2\pi\hbar/e$). Equi-gauge potentials are exactly radial emanating from their mother field line. The field lines are shown in their cross sections and point out of the plane. They are separated in space by some distance $d$. The equi-gauge potentials are clockwise numbered consecutively with the equi-gauge potentials of the right field line indicated by primes on the numbers. Since potentials add the superposition of the equi-potentials generates the dashed repulsive potential pattern in the space between the field lines, indicating that the interaction between a pair of parallel field lines is subject to repulsion.}}
\vspace{-0.3cm}\label{fig-equi}
\end{figure}

\subsection{\small{Gauge field geometry}}

Even though it is intuitive, the classical argument of the Lorentz force given before does thus not apply, at least not in the conventional form. From the fact that, for a single isolated field line, $\nabla\Lambda$ has only an azimuthal component \citep[see, e.g.,][]{aharonov1959,treu2011} with $\Lambda$ being proportional to the azimuthal angle $\theta$ given by 
\begin{equation}
\Lambda(\theta)= \frac{\Phi_0}{2\pi}\theta=\frac{\hbar}{e}\theta
\end{equation}
one concludes that the gauge field $\Lambda$ is constant in the radial direction -- meaning that radii are gauge-field `equi-potentials' as shown by the black radials emanating from the two circles representing field lines in Figure \ref{fig-equi}. 

Analytically one adds up the two gauge fields $\Lambda_1(\theta)$ of field line 1 and $\Lambda_{\,2}(\theta')$ of field line 2.  The angles $\theta, \theta'$ are measured in the respective proper frames of field line 1 and 2, the origin of the latter being displaced along the $x$-axis by the distance $d$ from the origin of the former. The total gauge field is a potential field which is additive, being the sum 
\begin{equation}\label{eq-sum}
\Lambda=\Lambda_1(\theta)\pm\Lambda_{\,2}(\theta')=\frac{\Phi_0}{2\pi}\big(\theta \pm\theta'\big)
\end{equation}
where the $+$-sign refers to parallel field lines, the $-$-sign to anti-parallel field lines. The angle $\theta'$ is to be transformed into the proper frame of field line 1 such that $\theta'(\theta,r;d)$ becomes a function of distance $d$ (in units $\lambda_m$) between the field lines, angle $\theta$ (in radians), and radius $r$ (also in units $\lambda_m$). The angle $\theta'$ maps to an angle $\theta$ via the relation
\begin{equation}
\tan\theta'=\frac{R\,\sin\theta}{R\,\cos\theta-1}, \qquad R=\frac{r}{d}
\end{equation}
which when used in the above sum yields the expression
\begin{equation}\label{pheq}
\Lambda(\theta, R)=\frac{\Phi_0}{2\pi}\left[\theta\mp\tan^{-1}\left(\frac{R\,\sin\theta}{1-R\,\cos\theta}\right)\right], \quad {d}>{\lambda_m}
\end{equation}
for the quantum-mechanically correct total gauge field in the space external to the two field lines.  The $R$ and $\theta$  dependence of the second term in the brackets in Eq. (\ref{pheq}) destroys the strictly radial pattern of equi-gauge potentials, with the main region of interest being $R<1$. 

The gauge field equi-potentials are obtained by holding expression (\ref{pheq}) constant. This yields the equi-gauge equation
\begin{eqnarray}\label{rad}
R(\theta,\tilde\Lambda)&=&\frac{\tan\big(\theta-\tilde\Lambda\big)}{\cos\theta\Big[\tan\big(\theta-\tilde\Lambda\big)\pm\tan\theta\Big]} \\
\tilde\Lambda&\equiv&\frac{2\pi}{\Phi_0}\Lambda = \mathrm{const}
\end{eqnarray}
Varying $\tilde\Lambda$ and calculating $R(\theta,\tilde\Lambda)$ for each fixed value of $\tilde\Lambda$ produces a pattern of equi-gauge potentials which now has become dependent of radius $R$. This dependence is enforced by the mere presence of another field line at distance $r=d$. Clearly, if the distance between the field lines $d\gg r$ is large, i.e. $R\ll 1$, this pattern degenerates to the original radial pattern of one isolated field line, as is seen from Eq. (\ref{pheq}). Again the $\pm$ signs apply to parallel and anti-parallel field lines.

\subsection{\small{Equi-gauge potential construction}}
\begin{figure}[t!]
\centerline{{\includegraphics[width=0.5\textwidth,clip=]{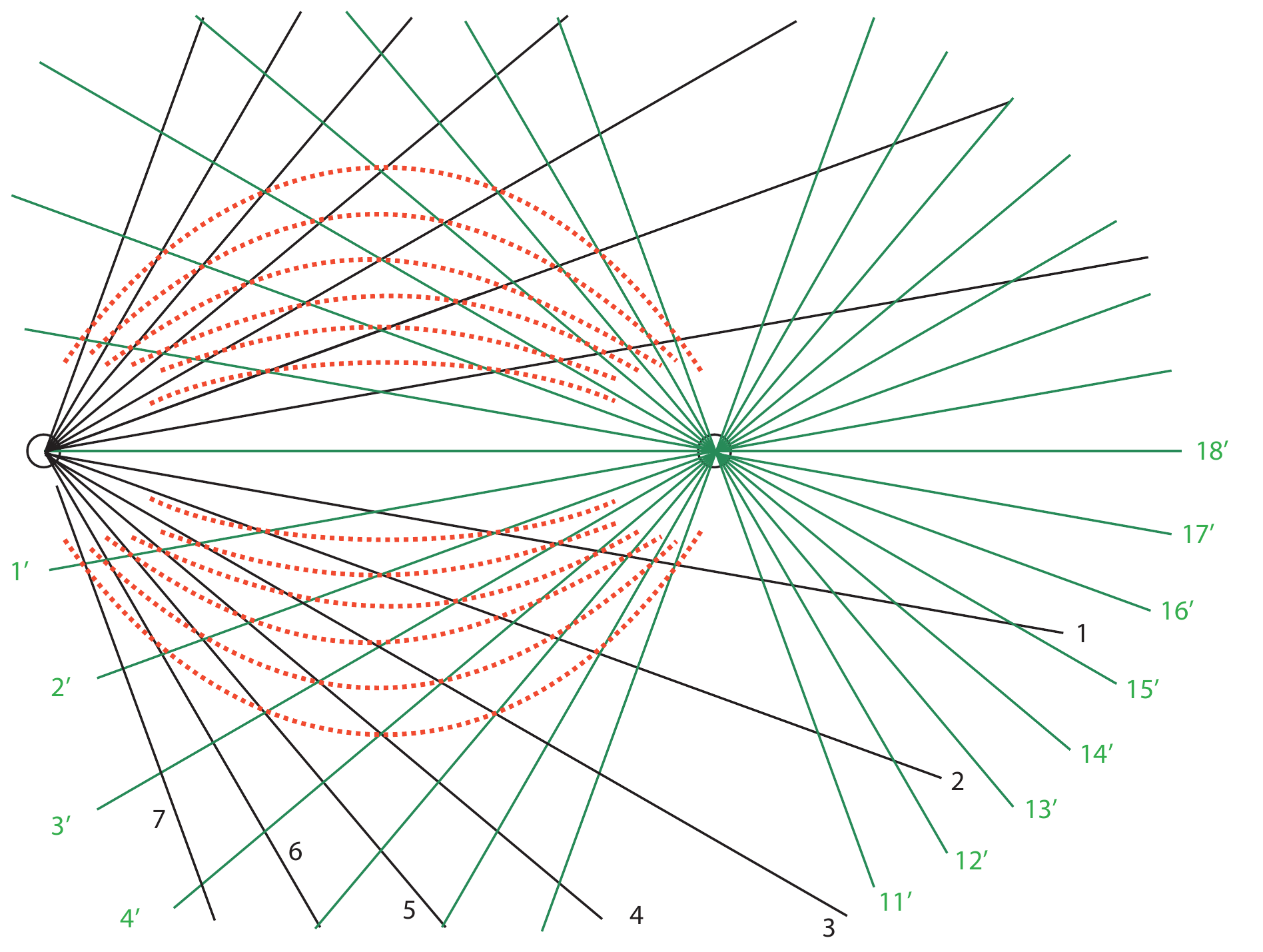}
}}
\caption[ ]
{\footnotesize {Superposition of the equi-gauge potential lines $\Lambda=$ const of two anti-parallel magnetic field lines. The field lines are shown in their cross sections with the left field line pointing out of the plane and the right field line pointing into the plane being spatially separated by some distance $d$. The equi-gauge potentials are again clockwise numbered consecutively. Because of the opposite direction of the field in the right field line the primed equi-gauge potentials are numbered anti-clockwise. Addition of the equi-gauge potentials yields the dashed equi-gauge superposition pattern of connected equi-gauge potentials in the region between the field lines. Such a pattern indicates attraction between the oppositely directed field lines mediated by the gauge fields.}}
\vspace{-0.3cm}\label{fig-anti}
\end{figure}

It is easy to geometrically construct the shape of the equi-gauge potentials by superposition. This has been done schematically in Figures \ref{fig-equi} and \ref{fig-anti} for the two respective cases of parallel and antiparallel field lines, where we plot the equi-gauge potentials for two (stretched) field lines in the perpendicular plane under the condition that each field line would be isolated in space and no other field lines would be present. In the parallel case the solitary patterns of both field lines are of course identical being numbered clockwise. In the antiparallel case they are numbered in opposite order (i.e.  the anti-parallel  field line radials are numbered anti-clockwise).  Superposing the two gauge fields produces the dashed curves  in these figures.

The important and intuitive observation is that for parallel field lines half way between the two field lines the superposition of the gauge fields creates a separation barrier in the gauge potential which forces the superimposed gauge field equi-potential field lines to deviate up to 90$^\circ$ from their radial directions. This enforces a pronounced radial dependence of the gauge equi-potential field according to Eq. (\ref{rad}). The pattern is similar to the equi-potentials produced by two electric charges of equal sign causing repulsion of the charges. Extrapolating to our case of two interacting parallel field lines we may conclude that it is the \emph{repulsive action of the gauge fields} between the two parallel magnetic flux tubes which keeps the parallel field lines on distance. It is this action which is responsible for the generation of the hexagonal structural lattice order of the field shown in Figure \ref{fig-hex}. 

Figure \ref{fig-anti} shows the plot of the equi-gauge potentials for the case when the magnetic fields are anti-parallel. In this case the left flux tube points out of the drawing plane, the right tube points into the plane.  By having turned the right flux tube by 180$^\circ$ into the plane, the rotational sense and thus the counting of the equi-gauge potentials is reversed. When superimposed with the equi-gauge potentials of the left flux tube, the picture of the dashed lines is obtained in this case. It is obvious that now the equi-gauge potentials of the two flux-tubes connect and an \emph{attractive} gauge-potential structure is obtained. 

Even though the physical implication of the repulsive  and attractive equi-gauge potentials is not quite clear in the ordinary quantum mechanical treatment given qualitatively here, we can conclude that the interaction between two field lines is mediated by the presence of gauge fields. Parallel magnetic field lines cause repulsive gauge field potentials, while anti-parallel field lines are subject to attractive  gauge field potentials. Clarification of the physical content awaits a treatment in terms of quantum electrodynamics --  i.e. the quantum electrodynamic solution of the Aharonov-Bohm problem with two flux tubes. Anticipating the solution, we boldly conclude from the electrodynamic analogy that the physical implication confirms the expectation that parallel field lines reject each other while anti-parallel field lines attract each other even though the space between the field lines is void of any magnetic fields. It is, however, filled with gauge fields which are responsible for the interaction.    

\subsection{\small{Numerous field lines}}
Equation (\ref{eq-sum}) can be generalised to many field lines by summing over their respective angles $\theta'_i$ and accounting for their varying distances from the origin $d_i$. Choosing one reference field line as the origin and the distance $d_0$ to one of its neighbours as direction of the $x$-axis, one has
\begin{equation}
\tan\,\theta'_i=\frac{R\sin\big(\theta+\alpha_i\big)}{R\cos\big(\theta+\alpha_i\big)-D_i}, \qquad R=\frac{r}{d_0},\ D_i=\frac{d_i}{d_0}>0
\end{equation}
Here, $0\leq\alpha_i\leq 2\pi$ is the angle the direction of $d_i$ makes with the direction of the $x$-axis, i.e. the direction of $d_0$. The normalised total gauge potential field is the sum over all contributions from the $i$ field lines
\begin{equation}
\tilde\Lambda(\theta,R)=\sum\limits_i \left\{\theta\mp\tan^{-1}\left[\frac{R\ \sin\big(\theta+\alpha_i\big)}{D_i-R\ \cos\big(\theta+\alpha_i\big)}\right]\right\}
\end{equation}
For some two-dimensional distribution of parallel $f_{\uparrow\!\!\!\uparrow}(D,\alpha)$ or antiparallel $f_{\uparrow\!\!\!\downarrow}(D,\alpha)$ field lines in space this expression becomes
\begin{eqnarray}
\tilde\Lambda(\theta,R)=\theta&-&\frac{1}{2\pi}\int\, \mathrm{d}D\ \mathrm{d}\alpha\  \big[ f_{\uparrow\!\!\!\uparrow}(D,\alpha) -f_{\uparrow\!\!\!\downarrow}(D,\alpha)\big] \times\nonumber\\
&\times&\tan^{-1}\left[\frac{R\ \sin\big(\theta+\alpha\big)}{D-R\ \cos\big(\theta+\alpha\big)}\right]
\end{eqnarray}
an expression which cannot be easily inverted for equi-gauge potentials. In a hexagonal lattice of lattice constant $d_0$ generated by many parallel field lines one has $D=n$ and $\alpha=\ell\pi/3$. Thus, 
\begin{equation}
f^\mathit{lattice}_{\uparrow\!\!\!\uparrow}(D,\alpha)\to 2\pi\delta(D-n)\delta(\alpha-\ell\pi/3)
\end{equation}
with $n\in\textsf{R}$ a natural number, $\ell=1, ..., 6$, and $f_{\uparrow\!\!\!\downarrow}(D,\alpha)=0$. The most important effect in this case is expected to result from nearest neighbours implying $n=1$. The gauge field pattern then repeats itself for any field line in the entire volume and is obtained from
\begin{equation}
\tilde\Lambda_\mathrm{nn}(\theta,R)=\theta - \sum\limits_{\ell=1}^{6} \tan^{-1}\left[\frac{R\ \sin\big(\theta+\ell\pi/3\big)}{1-R\ \cos\big(\theta+\ell\pi/3\big)}\right]
\end{equation}
Oblique field lines introduce further complications which we do not consider. On the other hand, importing antiparallel field lines will destroy the lattice locally causing lattice defects and topological reorganisation.

This theory is based on the notion of the additivity of the gauge potentials spanned by each of the field lines. As long as there is no other known interaction between magnetic flux quanta, superposition of the gauge fields is well justified. It will, however become distorted if some interaction potential has to be included. At the time being no such interaction potentials are know, at least to our knowledge.

\section{{Vacuum effects}}

A heuristic argument about the force between the flux tubes can be put forward as follows:
The gauge field 
$\nabla \Lambda= \mathbf{A}$
causes an electric potential 
\begin{equation}\label{eq-u}
U=-\partial\Lambda/\partial t 
\end{equation}
\citep[cf., e.g.,][pp. 220-223]{jackson1962} being of pure gauge nature. It is clear that the gauge field around an isolated field line is stationary, and $U=0$. In the presence of another field line, however, information is exchanged between the field lines, requiring time.  The gauge field becomes non-stationary, acquiring time dependence; the equivalent induced electrostatic potential is non-zero. 

The electrostatic energy the gauge field acquires in this case is obtained multiplying with charge $e$. This also defines a frequency $\omega_{\,\Lambda}$ via
\begin{equation}
eU=-e\frac{\partial\Lambda}{\partial t}=\mathrm{sgn}\left(\frac{\partial}{\partial t}\right)\hbar\omega_{\,\Lambda}\quad\longrightarrow\quad \omega_{\,\Lambda}=\frac{2\pi}{\Phi_0}\left|\frac{\partial\Lambda}{\partial t}\right|
\end{equation}
which suggests that the interaction between flux tubes is mediated by the exchange of massless particles -- photons -- of frequency given by the induced time derivative of the gauge field.

The time dependence of the gauge field in the Lorentz gauge is taken care of by the wave equation for $\Lambda$
\begin{equation}
\nabla^2\Lambda -\frac{1}{c^2}\frac{\partial^2\Lambda}{\partial t^2}=0
\end{equation}
of which the solution $\Lambda$ is subject to the boundary conditions on the surfaces of the two flux tubes. These prescribe that $\nabla\Lambda=\mathbf{A}$ on both surfaces. 

Formally the potential caused by the gauge field gives rise to a gauge-Coulomb force 
\begin{equation}
\mathbf{F}=e\nabla U= -e\frac{\partial\nabla \Lambda}{\partial t}=-\frac{2\pi\hbar}{\Phi_0}\frac{\partial\nabla \Lambda(\theta,r,t)}{\partial t} 
\end{equation}
which in the presence of another field line evolves a radial component the sign of which depends on the mutual orientation of the field lines.  Formally, field lines behave like electric charges of value $2\pi\hbar/\Phi_0$. The force $\mathbf{F}=\mathrm{d}\mathbf{p}/\mathrm{d}t$ is the time derivative of a momentum $\mathbf{p}=\hbar\mathbf{k}$. However, there are no massive charged particles involved on which the force could act in the empty space between the field lines. Hence the change in momentum
\begin{equation}
\Delta\mathbf{p}= -\frac{2\pi\hbar}{\Phi_0}\nabla\Lambda(\theta,r)
\end{equation}
must be experienced by the flux tubes only, where for two field lines $\Lambda(\theta,r)$ is given in Eq. (\ref{pheq}). $\Delta\mathbf{p}$  causes acceleration and displacement of the field line in the presence of another field line at distance $d$. 

\subsection{\small{Virtual pairs}}

The problem consists in understanding how, in the absence of any massive charged particles and the mere presence of gauge fields, the force between two separate flux tubes is transmitted across the field-free and matter-free space between field lines. 
The only possibility is the inclusion of the vacuum as an active medium. Field line interaction will be understood only when referring to low-energy vacuum theory. 

The energy carried by the gauge field is of order 
\begin{equation}
eU\sim h\frac{\partial \theta}{\partial t}\sim 10^{-15}\dot\theta\quad\mathrm{eV}
\end{equation}
which is small. Referring to  Dirac vacuum with all negative energy states filled by Fermions, spontaneous pair creation of \emph{real} particles is impossible as it requires energies $\gtrsim 1$ MeV, or $\dot\theta\sim10^{\,21}$ Hz. One, however, with $\Delta\epsilon=2m_ec^2$ the energy of an electron-positron pair, observes from the energy-time uncertainty relation $\Delta\epsilon\Delta t\sim h$ that this frequency corresponds to a time uncertainty 
\begin{equation}\label{eq-deltat}
\Delta t \sim \pi \hbar/m_ec^2 \sim 10^{-21}\quad\mathrm{s}
\end{equation}
which allows for the creation of virtual pairs, living on `borrowed time'. Hence, the region of the gauge field gradient between the two flux tubes is filled with a cloud of \emph{pairs of virtual electrons and positrons} each of them present for a time $\Delta t$ and, before disappearing and being replaced by another pair, each propagating a Compton wavelength $c\Delta t \sim h/m_ec\sim 10^{-12}$ m in the equivalent gauge-electric field $-\nabla U$ in opposite directions, causing screening currents. This is the vacuum polarisation effect, well known from Quantum Electrodynamics \citep[cf., e.g.,][and any other advanced text on QED]{aitch1993,kaku1993,landau1997}. 

Here its effect is to screen the equivalent gauge electric field and to reduce the bending of the gauge equi-potentials in  order to restore straight radial gauge field lines of independent magnetic flux tubes, either pushing the field lines some distance apart or causing attraction and annihilation of the antiparallel flux, depending on the mutual directions of the field lines. In this way the force is transmitted to the magnetic field line by the presence of the virtual particle cloud which exists only in the region of $\nabla\Lambda\neq 0$ and thus only when another field line is added. One realises that this is a dynamical and thus time dependent process. It ceases immediately when the field lines get sufficiently far apart from each other and the radial dependence of the inter-field line gauge field disappears. Otherwise, when the field is confined from the outside, the field lines will be in dynamical equilibrium with the confining force, being continuously surrounded by a cloud of virtual pairs.

\subsection{\small{Vacuum topology}}

\begin{figure}[t!]
\centerline{{\includegraphics[width=0.35\textwidth,clip=]{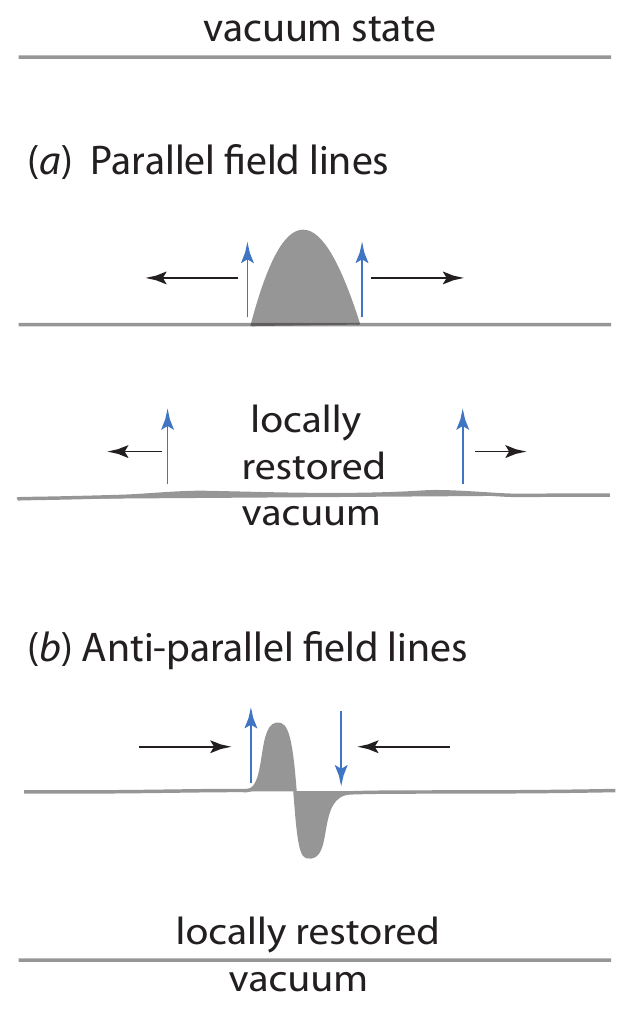}
}}
\caption[ ]
{\footnotesize {Topological distortion of vacuum by two field lines. ($a$) Two parallel field lines (blue $\uparrow\uparrow$) placed at small distance with their gauge fields causing distortion of vacuum in space between field lines. The vacuum restores its state locally by pushing the two field lines apart (black arrows) in opposite directions. ($b$) Two anti-parallel field lines (blue $\uparrow\downarrow$) causing a sinusoidal distortion. The vacuum state is restored by attraction when the field lines annihilate over some parallel part locally. }}\label{fig-vac}
\vspace{-0.3cm}
\end{figure}

Modern field theory gives a topological interpretation of the physical vacuum as the (average) minimum-energy ground state of an interacting many-body system which is a symmetrical equilibrium. In our microscopic case we may assume a flat vacuum equilibrium state on the scale of the field line flux tubes. Putting one field line into vacuum distorts it locally adding the cylindrically symmetrical gauge potential which does not do any serious harm to the vacuum equilibrium state because it lacks any radial and time variations. The situation is two-dimensional only. With two and more field lines, however, the situation becomes different as shown in Figure \ref{fig-vac}. 

When two parallel field lines are put into the vacuum close to each other, the distortion of the vacuum by the gauge potential causes humps in the vacuum field. These are the result of the creation of clouds of virtual electron positron pairs in the regions of finite gauge potential gradients and the virtual particle fields and currents involved. The vacuum needs -- and acts -- to restore the flat ground state. This is achieved by stretching the vacuum potential and pushing the humps apart. Solving the quantum electrodynamical problem of this interaction is a formidable task. An attempt in this direction is done below. 

For a qualitative argument we refer to Fig. \ref{fig-equi} which shows that the bending of gauge-field equipotentials is strongest near the two anti-parallel field lines close to the straight line connecting their centres. The gradients are perpendicular to this line and are parallel for the two field lines. Hence, the clouds of virtual particles concentrate here and carry parallel virtual currents which interact repulsively. It is thus the repulsive force between the virtual current carried by the virtual particles generated in the vacuum by the two interacting field lines which exerts a force on the field lines. Under the action of this force the two field lines separate in space, the gauge potentials stretch radially out and the virtual particles ultimately disappear. This action may be interpreted as attributing a virtual (time-dependent) mass $M_{\Phi_0}$ to the field lines, which is the total mass of the cloud of virtual particles 
\begin{equation}
M_{\Phi_0}(t)= 2m_e\int d^3 x d^3p f_{virt}(\mathbf{x},\mathbf{p},t)
\end{equation}
where $2m_e$ is the mass of the virtual electron-positron pairs, and the integration is taken over the volume of non-vanishing gradient of the gauge potential field. The function $f_{virt}(\mathbf{x},\mathbf{p})$ is the properly normalised distribution function of the virtual particles with momentum $\mathbf{p}$ at location $\mathbf{x}$. The inertia of the virtual particle cloud thus attributes an inertia to the field line and mediates a repulsive force acting between the field lines. 

Similarly, if two anti-parallel field lines are put into the vacuum, the vacuum assumes a sinusoidal distortion (as is schematically shown in the lower part of Fig. \ref{fig-vac}) which can be most simply relaxed by attracting and annihilating. 

\subsection{\small{Briefing on the quantum electrodynamic approach}}
Calculation of this mechanism requires the full technique of a distorted quantum electrodynamic vacuum theory. This will not be explicated here in detail. It has, for strong electromagnetic fields below the critical electric field $|\mathbf{E}|< E_\mathit{crit}=m_e^2c^3/e\hbar\approx 2.2\times10^{17}$ V/m, originally been given by \citet{euler1936}. [The corresponding critical magnetic fields have strengths $B_\mathit{crit}\approx 4.4\times10^{17}$ T.] Its quantum-field theoretical form was developed by \citet{schwinger1951}.\footnote{For later refinements and application see, e.g., \citet{landau1997} and \citet{itz1980}. For a modern recount and application to very strong fields in pulsars and magnetars, showing that vacuum polarisation effects relax the condition on the upper limit on the magnetic field strength allowing for the existence of magnetars, see  \citet{heyl1997a,heyl1997b}.}  Here we sketch the mechanism in view of application to our problem without going into the details of its complicated mathematics. 

As noted above, the external gauge potential of two (or more) isolated magnetic flux tubes causes an equivalent electric field by producing spatial gauge field gradients.\footnote{Recall that one single flux tube does not give rise to such effects if assuming that, locally, it can be considered an infinitely extended string. Of course, magnetic fields have no divergence, and therefore, the flux tube or field line must at some location become bent and return either into itself to close or end up on an external source. Both will cause distortion of the vacuum at some place, which, however, locally is not felt if displaced sufficiently far.} These correspond to a local electric field which necessarily polarises the vacuum, even in our case of very weak fields. The production rates of real pairs \citep{euler1936,heyl1997a} are low in weak fields and increase exponentially with increasing field. The magnetic and electric fields which we are interested in are much less than the critical fields, $|\mathbf{B}|\ll B_\mathit{crit}, |\mathbf{E}|\ll E_\mathit{crit}$, where $\mathbf{E}$ is the field caused by the external gauge potential $\Lambda$. Clearly, in this case no real pairs can be generated. This has been explicated above several times already.  Then the required vacuum polarisation is produced by creation of \emph{virtual} electron-positron pairs living on the short `borrowed' time (borrowed from quantum uncertainty), though being continuously reproduced again and again and thus contribute to a quasi-permanent cloud of virtual pairs. 

\subsection{\small{Virtual pair production rate}}
Defining $\zeta=|\mathbf{E}(r,\theta)|/E_\mathit{crit}$, we can estimate the pair-density production rate $\kappa(\zeta)$ by referring to one of the above papers, where the problem is solved for real pairs in quantum electrodynamics. There, it has been shown \citep{euler1936,schwinger1951,landau1997,itz1980} that the pair-density production rate out of the vacuum in the presence of an electromagnetic field is defined as being proportional to the imaginary part
\begin{equation}
\kappa(\zeta)=\frac{\mathrm{Im}\ {\cal L}}{2\pi\hbar}
\end{equation}
of the (complex) interaction Lagrangean $\mathcal{L}= \mathcal{L}_0+\mathcal{L}'$ of the electromagnetic field with the vacuum.  $\mathcal{L}$ can be expressed through the electrodynamic Lorentz invariants 
\begin{eqnarray}
\mathcal{I}&=&F_{\mu\nu}F^{\mu\nu}\equiv 2\left(\frac{|\mathbf{B}|^2}{2\mu_0}-\frac{\epsilon_0|\mathbf{E}|^2}{2}\right), \\ \mathcal{K}&=&\in^{\lambda\rho\mu\nu} F_{\lambda\rho}F_{\mu\nu}\propto \mathbf{E\cdot B}
\end{eqnarray}
where $F_{\mu\nu}$ is the covariant electromagnetic field tensor. In our case, where no magnetic field exists outside the flux tube, the second invariant, which is the magnetic field-aligned electric component, vanishes identically, yielding for the Lagrangean
\begin{equation}
\mathcal{L}\big(\mathcal{I},\mathcal{K}\big)=-{\textstyle\frac{1}{4}}\mathcal{I}
\end{equation}
Calculating all these functions and solving for the imaginary part of the Lagrangean is possible in the two limits of large and small electric field ratios $\zeta$. In our problem we are interested only in the very small ratio limit $\zeta\ll\!\!\!\!\!\!\!< 1$. The calculation is lengthy and subtle. We give here the main steps only.

The expression for $\kappa$ simplifies substantially in the small $\zeta$ case, though $\mathcal{L}'\big(\mathcal{I},\mathcal{K}\big)$ is still a rather complicated integral expression. For small $\zeta$, the pair production rate is to be taken at
\begin{equation}
\kappa(\zeta)=\frac{1}{2\pi\hbar}{\cal L}\Big(\mathcal{I}=-2\zeta^2;\,\mathcal{K}=0\Big)
\end{equation}
This can be expressed \citep{heyl1997a,heyl1997b} as
\begin{eqnarray}
\kappa\approx C\Big\{{\textstyle\frac{1}{2}}\pi&+&2\zeta \Big[2\mathrm{Re}\mathcal{J}(\zeta)-\zeta\big( \ln\,\zeta+\ln\ 4\pi\ +1\big)\Big] - \nonumber\\[0.5ex]
&-&\zeta^2\Big[{\textstyle\frac{1}{3}}\pi+8\ \mathrm{Im}\mathcal{J}(\zeta)\Big]\Big\}
\end{eqnarray}
with $\mathcal{J}(\zeta)$ defined as the integral
\begin{equation}
\mathcal{J}(\zeta)\equiv\int\limits_0^1\mathrm{d}s\ \ln\left\{ \Gamma\Big[1+s\big(\frac{i}{2\zeta}-1\big)\Big]\right\}
\end{equation}
Since, in our case, $\zeta\ll\!\!\!\!\!\!\!\!< 1$ is a very small number, we can use the asymptotic expansion 
\begin{equation}
\Gamma(az+b)\sim \sqrt{2\pi}(az)^{az+b-\frac{1}{2}}\mathrm{e}^{-az}, \quad (a>0,\ |\mathrm{arg}|\,z<\pi)
\end{equation}
for the $\Gamma$-function of complex argument. Taking the logarithm, integrating each term and carefully watching to take the lower limit of the integral to avoid divergence, we find for the real and imaginary parts of the integral
\begin{eqnarray}
\mathrm{Re}\mathcal{J}&\sim & -\frac{1}{2}\ \ln\left(1+\frac{1}{4\zeta^2}\right) \\
\mathrm{Im}\mathcal{J}&\sim& -\frac{1}{4\zeta}\left[1-\ln\ \left(1+\frac{1}{4\zeta^2}\right)\right] -\arctan\frac{1}{2\zeta}
\end{eqnarray}
Inserting into the expression for the pair production rate we finally find for $\kappa(\zeta)$, up to second order in $\zeta$,
\begin{equation}
\kappa(\zeta)\sim C \zeta\Big(1-\frac{\pi}{6}\zeta\Big), \qquad \zeta\ll1
\end{equation}
At these low values of $\zeta$, the pair-density production rate is a linear function of $\zeta$. The proportionality factor, 
\begin{equation}
C\equiv m_e^2c^3/8\pi^2\hbar^4 = 4\times 10^{57}\quad \mathrm{m}^{-3}\, \mathrm{s}^{-1} 
\end{equation}
is a huge number \citep[cf., e.g.,][]{heyl1997a}, indicating that in stronger electric fields the production of pairs, under conditions when the quantum electrodynamic theory is applicable, is quite efficient. However, in our case, $\zeta$ is extremely small, thus $\kappa$ will be substantially reduced. 

This is, however, not yet the full story. Since no real pairs can be produced by the very small expected electric field strengths, $|\mathbf{E}|$, which result from the presence of the deformed gauge potential $\Lambda$, it makes no sense to ask for the pair  production rate. What is of interest, is the density of virtual pairs which have been generated at the end of the `borrowed' time, $\Delta t$, from uncertainty,  Eq. (\ref{eq-deltat}), as this will be the upper limit of virtual pairs which the equivalent electric field that is generated by the gauge potential can afford. This number is obtained from the definition of $\kappa(\zeta)=\mathrm{d}N_\mathit{pairs}/\mathrm{d}t$ as
\begin{equation}
N_\mathit{pairs}=\int\limits_0^{\Delta t}{\mathrm{d}t} \kappa(\zeta)
\end{equation}
$\kappa(\zeta)$ contains the electric field, which is given from Eq. (\ref{eq-u}) by the gradient of $U$ as the time derivative of the gauge potential $\Lambda$. Making use of this property, the integral can be done, yielding for the number density of virtual pairs
\begin{equation}
N_\mathit{pairs}\approx\frac{C}{E_\mathit{crit}}\Big|\nabla\big[\Lambda(\Delta t, r, \theta)-\Lambda(0, r, \theta)\big]\Big|
\end{equation}
Unfortunately, this form cannot be treated further. In order to obtain an absolute upper limit of the virtual pair density, we simply multiply $\kappa(\zeta)$ by $\Delta t$, finding
\begin{equation}
N_\mathit{pairs}\lesssim 4\times 10^{36}\zeta
\end{equation}
This also yields a limit on the mass density of the pair cloud $m_{\Phi_0}\sim 2m_eN_\mathit{pairs}$, which is the mass density attached to the magnetic flux tubes. 

Estimating a reasonable value for $\zeta$ is difficult. Observations in space suggest that the electric field related to a single field line must be very small, much less indeed than any reasonable field which the E$\times$B convection velocity of magnetic plasma fluctuations would suggest. Convection electric fields in space are of the order of $\sim$ mV/m, while lower limits on the electric fluctuations range around $10^{-9}$ V/m. In order to be on the safe side when having in mind that we want to apply this theory to reconnection in magnetic fields of the order of O(10) nT,  we boldly assume that $|\mathbf{E}|\lesssim 10^{-15}$ V/m. This still yields a virtual pair density of 
\begin{equation}
N_\mathit{pairs}\sim 10^3\quad \mathrm{m}^{-3} 
\end{equation}
for the plasma sheet in the magnetospheric tail, corresponding to $\sim 10^{-3}$ cm$^{-3}$ which is well below any observed plasma density in this region. These pairs are located only in a small region close to each of the magnetic field lines, however, mediating the field line dynamics on the inter-field line scale. Any average density of such virtual pairs will be even much less when averaged over the volume.

The electric current density carried by any such cloud of virtual pairs is also small. It can be estimated as
\begin{equation}
\mathbf{J}_\mathit{pairs}\approx 2e N_\mathit{pairs} V_\mathit{pairs}
\end{equation}
where the velocity is given by $V_\mathit{pairs}\approx E\Delta t$, since the pairs become accelerated in the equivalent gauge electric field only for the borrowed time $\Delta t$. As expected, this current density is small, being of the order of 
\begin{equation}
\mathbf{J}_\mathit{pairs}\sim 10^{-58}\quad \mathrm{A\ m}^{-2} 
\end{equation}
in the vicinity of one field line. The presence of the current gives rise to a small Lorentz force density
\begin{equation}
\mathbf{F}_\mathit{L}\sim \mathbf{J}_\mathit{pairs}\times\mathbf{B}\sim 10^{-66}\quad \mathrm{N\ m}^{-2} 
\end{equation}
in a field of $B\sim 10$ nT. This force is acting on the field line via pushing the cloud of virtual particles and mass density 
\begin{equation}
m_\mathit{pairs}\sim 10^{-27}\quad \mathrm{kg\ m}^{-3} 
\end{equation}
around. Interestingly, the mass density of such a cloud of virtual pairs corresponds to just about one proton per m$^3$. This is the virtual mass attributed to the magnetic flux tube (or field line) generated in the quantum electrodynamic process of distorting and polarising the vacuum. 

\section{Implications for collisionless reconnection}
The sub-microscale quantum dynamics of field lines has interesting implications for the mechanism of collisionless reconnection the ultimate cause of which has so far remained in the dark. Observations as well as numerical simulations unravelled many of the macro-scale properties of reconnection during the past few decades.

It was proposed that reconnection takes place in the so-called `ion diffusion region' (though there is not any remarkable diffusion present here), which is the volume of ion-inertial radius $r<\lambda_i=c/\omega_i$ surrounding the reconnection \textsf{X}-point ($\omega_i$ is the ion plasma frequency). Ions become non-magnetic here, decouple from the magnetic field $\mathbf{B}$, follow their inertia and, in addition, become accelerated in any electric field that would be present, e.g. the cross-tail convection electric field in the case of the magnetospheric tail. No convincing mechanism has been identified being capable of causing reconnection on the scale of the 'ion diffusion region'. 
 Electrons remain magnetised, freeze the magnetic field to their cyclotron orbits and transport it convectively across the `ion diffusion region'. In this way they cause the Hall currents proposed by \citet{sonnerup1979}. 
 
The physics of reconnection is thereby deferred to the electron-inertial scale region $r<\lambda_e=c/\omega_e$ around the \textsf{X} point (with $\omega_e$ the electron plasma frequency) where the electrons become as well non-magnetic. This, however, implies that the magnetic field would be transported -- in some way -- into the very centre of the current sheet in order to get into close contact with the oppositely directed field and magnetic flux. 
\begin{figure*}[t!]
\centerline{{\includegraphics[width=0.85\textwidth,clip=]{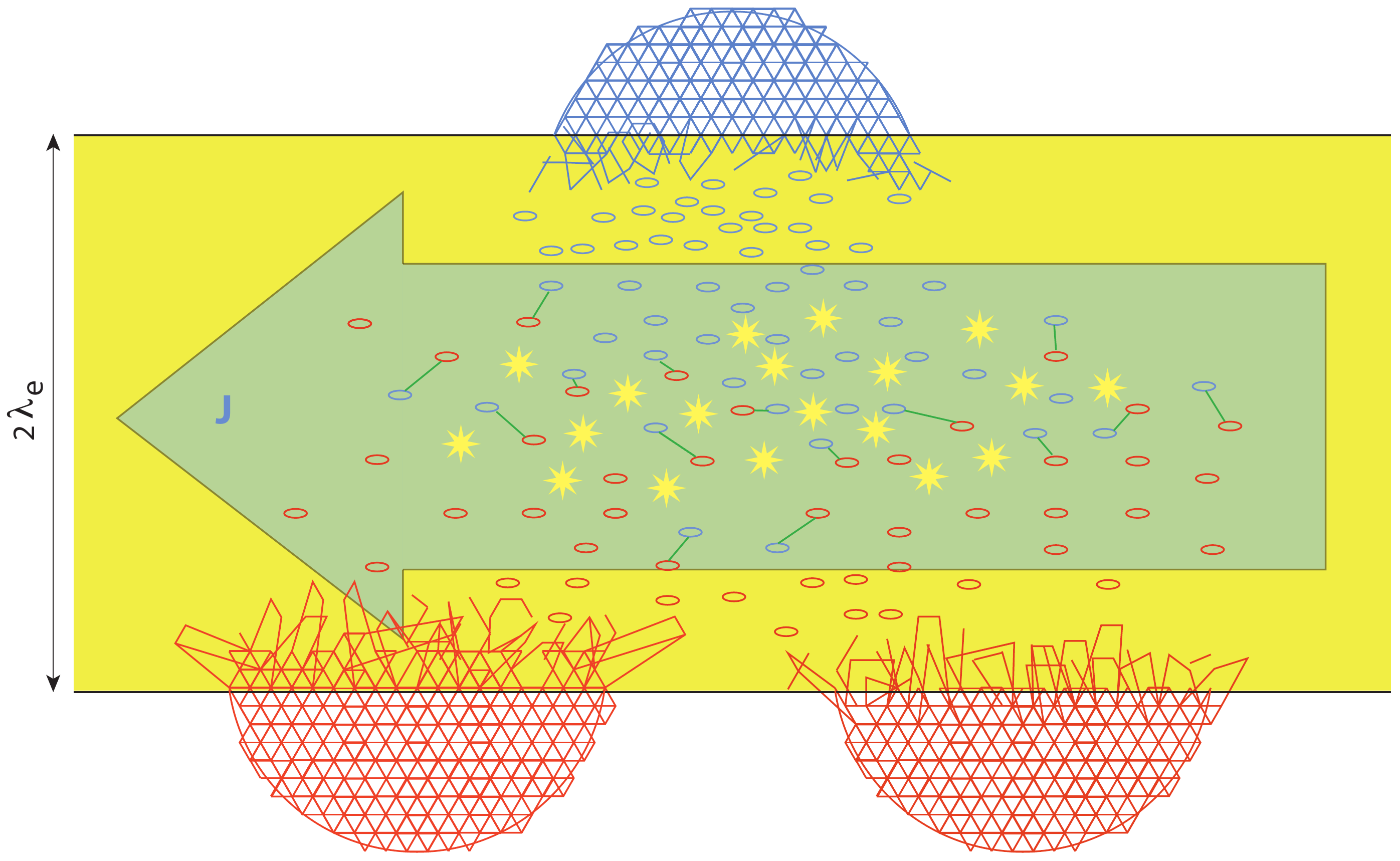}
}}
\caption[ ]
{\footnotesize {Three meso-scale flux tubes encountering the electron inertial region at the centre of a (symmetric) current layer which separates two plasmas with antiparallel fields. The (blue) field above the current $\mathbf{J}$ points into the plane, the (red) field below the current points out of the plane. Above and below the electron inertial layer the field is frozen to the electrons (not shown) forming a lattice structure. When entering the electron inertial region, the lattices break off and dissolve as the fields become released from the frozen-in state. Under the action of the repulsive  gauge fields they seek to achieve larger distances whereby entering the central region. Meeting field lines of different polarity from the other side, they feel  attractive gauge forces (indicated by green lines), approach each other and annihilate (yellow stars). The distribution of merging centres is statistical. Merged field lines relax and join up to produce macro-scale reconnection effects: jets etc.}}\vspace{-0.3cm}\label{fig-diss}
\end{figure*}

Classically it is by no means obvious how this transport could be realised in the collisionless case on the small scale. Once the electrons become non-magnetic the magnetic fields become independent of the plasma on the scale $r<\lambda_e$. Since the magnetic field has no inertia, any bending of the field causes relaxation and stretching of the field lines. The massless and elastic field lines will snap back to the inflow region \citep[as suggested by][]{baum2010}. Further classical inward transport requires an enhanced magnetic pressure at the boundary of the electron inertial region or some other mysterious cause. The former is possibly only under driven reconnection conditions when the plasma is compressed from the outside by some brute force, a situation which is realised at the magnetopause under conditions of solar wind impact. Macro-scale consequences of such driving have been investigated by \citet{pritchett2005} with the help of sophisticated and carefully performed numerical simulations.   

In the absence of driving, to which most reconnection models and simulations refer, the current sheet will remain free of any magnetic fields. Simulations circumvent this case usually by either imposing, globally or locally, an artificial resistivity -- or any other kind of dissipative mechanism -- in the current sheet \citep[implicitly referring to the respective original models of][]{parker1958,petschek1964} or simply impose brute force seed \textsf{X}-points in order to initiate reconnection \citep[the method described in][and widely used by the PIC community]{zeiler2002}. All these attempts do not explain the onset of reconnection as a fundamental physical process. They, instead, properly account for the various macro-scale effects of reconnection which may occur under various conditions when reconnection has already set on and continues to take place.\footnote{For a conservative review, see \citet{biskamp2000}. A recent critical though very concise account of the available reconnection models is given in \citet{baumjohann2012}.}   This approach is well justified under the assumption that some unspecified mechanism generates the seed \textsf{X} points. Such mechanisms have been based on magnetic fluctuations or the action of some electromagnetic instability \citep[e.g. whistlers, kinetic Alfv\'en waves, or  the Weibel modes proposed in][]{treumann2010}.

Starting from the field-line concept, we investigated the sub-microscale merging process between two isolated field lines \citep{treu2011} assuming that two antiparallel field lines were brought into mutual contact. No argument was provided of how anti-parallel field lines could enter the electron inertial range. Based on the theory of field-line interaction developed in this paper, we have arrived at the position to complete the reconnection picture. 

In the collisionless case the magnetic field remains frozen to the electron cyclotron orbit outside the electron inertial scale region. E$\times$B drift transports the flux tube frozen to the electron cyclotron cross section toward the centre of the neutral current layer. The slowly weakening Harris-sheet magnetic field experienced during this motion lets the electron-cyclotron flux tube expand as the square of the electron gyroradius $r_\mathit{ce}\sim \sqrt{T_\perp/B^2}\sim1/\sqrt{B}$, where we have taken into account the adiabatic increase of the  perpendicular temperature $T_\perp\sim B$. This causes the lattice constant $d$ of the frozen field lines to increase at the same rate under the action of the weakening field and repulsive action of the gauge potentials.

Ultimately, the electrons approach the boundary of the electron inertial region near the centre of the current sheet and demagnetise (see Figure \ref{fig-diss}). Their gyroradius has increased to become larger than the electron inertial scale, here. At this instant, the lattice of frozen magnetic field lines explosively dews, field lines desert from adiabatic electron slavery, becoming released, and the lattice structure dissolves \citep[undergoing phase transition similar to two-dimensional lattice melting in solid state physics, cf., e.g.,][]{huang1987}. 

From now on, the parallel field line dynamics is determined by the stresses resulting from field line bending and repulsive forces. The least bent field lines continue moving and enter deeper into the neutral sheet current layer, as their restoring stresses are smaller than the repulsive gauge forces of the denser packed companion field lines in their backyard plasma. For them the repulsive inter-field line gauge fields dominate their dynamics pushing them ahead to penetrate the neutral current layer. Here, they meet field lines of opposite direction which entered the neutral current layer from the other side of the current by the same mechanism. The oppositely directed elementary flux tubes experience the attraction in the mutual gauge fields and become accelerated toward each other in order to approach quickly and, when coming into close contact, collide and annihilate the flux quanta stored in them over a certain parallel length $\ell_\|$. This elementary merging process has been investigated in detail \citep{treu2011} and will not be repeated. 

Since many a number of field lines are added to the neutral layer, it is clear that a large number of field lines participate in merging and annihilation, adding up to macro-scale reconnection-flux tubes and causing the different inferred macro-scale effects reconnection offers under the various external plasma conditions. In this way, the dynamical interaction of field lines provides the wanted consistent picture of spontaneous onset of reconnection. 

In view of a possible observational confirmation of the sub-microscale merging of field lines and cause of reconnection one is currently bound to spacecraft measurements which, unfortunately, cannot resolve any of the sub-microscales under question. Indirect evidence is the only way of experimentally checking the reality of our theory. This evidence may be given by observation of wave or radiation processes in the three different stages of merging in the chain of reconnection:  initial single field line merging, followed by inclusion of dielectric effects, mass loading by electrons, and finally the already known macroscopic stage of mass loading by ions. The first two interesting stages are sub-microscale \citep[][]{treu2011}. They occur when the curvature radius $r_c$ of the merged field lines is shorter than the Debye length, 
\begin{equation}
\lambda_m < r_c < \lambda_D
\end{equation}
As long as this holds, the merged and strongly kinked field line relaxes like in vacuum. This field line relaxation is identified as free-space electromagnetic radiation of frequency $f_\mathit{em}$, i.e. high-frequency electromagnetic radiation in vacuum. Since it is expected that in reconnection in a current layer very many field lines merge, this will cause an electromagnetic radiation spectrum in a fairly broad frequency range
\begin{equation}
c/\lambda_D < f_\mathit{em} < c/\lambda _m
\end{equation}
Rewriting this expression yields
\begin{equation}
\frac{c}{\beta_e\lambda_e} < f_\mathit{em} < \left(10^2 - 10^3\right) \left(B_\mathrm{nT}\right)^\frac{1}{2}\quad \mathrm{GHz}
\end{equation}
where the lower limit is determined by the velocity ratio $\beta_e=v_e/c$  and the electron inertial scale length $\lambda_e=c/\omega_e$, and $B_\mathrm{nT}$ is in nT. This can also be written as
\begin{equation}
\frac{30\omega_e}{\sqrt{T_e}} < f_\mathit{em} < \left(10^2 - 10^3\right) \left(B_\mathrm{nT}\right)^\frac{1}{2}\quad \mathrm{GHz}
\end{equation}
where $T_e$ is in units of $10^{-3}m_ec^2= 511$ eV. This indicates that the emission band is quite far above the local plasma frequency. Each of the merged field lines contributes to it by emitting a \emph{falling tone} in the electromagnetic spectrum. 

Once, however, the curvature radius reaches the Debye length, dispersion of the electromagnetic waves change due to modification of the dielectric properties, and the emitted radiation becomes cut off at the plasma frequency. The short wavelength high frequency radiation escapes from the reconnection site and enters the surrounding magnetised plasma where its polarisation properties come into play. One may expect that the original emission would be non-polarised, i.e. emission is isotropic. Hence, in the magnetised plasma it will split to equal amounts into right and left-hand polarisations. This produces a mixture of high frequency O-mode and X-mode radiation. 

So far it is not clear yet whether or not the initial radiation will indeed be isotropic. The degree of polarisation will depend on the mechanism of merging of single field lines which will to some degree also be affected by the torsion of the magnetic field line flux tube, and this torsion should determine the polarisation properties of the emitted radiation. Possibly the torsions, and thus the polarisations,  are different on both sides of the current sheet, however, which would make a distinction for emission of radiation on either side of the reconnection site. The related questions are still open to investigation.

At later stages when the curvature radius exceeds the electron inertial scale, the electrons become magnetised, and the fluctuations propagate in the whistler or Z-mode bands. Observation of the form of radiation in relation to reconnection signals the sub-microscopic mechanism of field line merging.

Reconnection itself is identified to be a sub-microscale phenomenon. It can be understood on the basis of the quantum concept of magnetic field lines and their interaction via their external gauge fields. Macroscopic reconnection then becomes an intrinsically three-dimensional phenomenon:  the interaction of many merging (sub-microscale) antiparallel field lines forming large numbers of reconnection specks. As a by-product, it also makes clear why spontaneous reconnection is a statistical phenomenon which occurs in small patches resembling turbulent (or patchy) reconnection.

\conclusions
The present communication does not resolve all the problems related to the dynamics of magnetic field lines and even less those related to the problem of (macro-scale) reconnection. It, however, suggests that the ultimate understanding of the microphysics of reconnection probably requires inclusion of magnetic field line dynamics, i.e. the dynamics of magnetic flux quanta $\Phi_0$ which in a given magnetic field $\mathbf{B}$ are enclosed into extended lines of well defined cross section $\pi\lambda_m^2$. Merging and annihilation of such flux quanta is possible only by direct contact of anti-parallel sections of the field-line flux tubes.

This can be understood as being due to the interaction of field lines via their external gauge fields. Most of the space between field lines is void of any magnetic fields. The interaction in question can be repulsive or attractive. Repulsive interaction is found between parallel field lines and, if the magnetic field is confined to a certain spatial volume of finite area, causes a hexagonal lattice structure of the field. On the other hand, attractive interaction occurs between anti-parallel field lines. The combination of repulsion and attraction is the main reason why field lines can enter into the centre of a neutral current layer separating plasmas of opposite magnetic field direction and is thus the basic cause of sub-microscale field line merging. It, in principle, solves the problem of reconnection on the quantum level. 

The discussion given in this communication is mostly qualitative (or semi-quantitative). However, it clarifies the main physics with only a limited amount of reference to the full quantum electrodynamics instrumentation of magnetic field line dynamics. We have, however, demonstrated that the gauge field external to a magnetic field line is capable of creating a small number of virtual electron-positron pairs which live on the borrowed time of quantum mechanical uncertainty. These pairs contribute to a very small though finite current density in the region between adjacent flux tubes and cause a weak Lorentz force on the pair cloud and elementary magnetic flux tube. This is the basic quantum physics of field line interaction. 

Macroscopically observed reconnection effects come to birth when many a number of field lines become involved, merge, relax and become ultimately mass loaded. The resulting chain of processes for two anti-parallel field lines has been described in an earlier paper \citep{treu2011}. 
Involvement of a very large number of merging field lines in some particular location requires a proper statistical approach. 

The macroscopic effects of reconnection will be rather different for different external conditions in the interacting plasmas and different parameter settings. However, the sub-microscale cause of merging and reconnection is quite general and independent of the macro-scale settings. It involves attracting magnetic field lines of opposite direction in order to penetrate into the centre of the neutral current layer and for coming into mutual contact, a problem which has been treated in the present communication. Under collisionless conditions, this kind of attraction is independent of any external macro-scale differences.

\begin{acknowledgements}
This research was part of an occasional Visiting Scientist Programme in 2006/2007 at ISSI, Bern. RT thanks the ISSI librarians, Andrea Fischer and Irmela Schweizer, for their assistance, and  Andr\'e Balogh, former Director at ISSI, for his continuous encouragement. He acknowledges the head-shaking ISSI directorship of hesitantly tolerating work on `exotic' problems. He appreciates the friendship of Johannes Geiss, Founder and Director emeritus of ISSI.
\end{acknowledgements}

\end{document}